# Formation and Microwave Losses of Hydrides in Superconducting Niobium Thin Films Resulting from Fluoride Chemical Processing

Carlos G. Torres-Castanedo,[†,1] Dominic P. Goronzy,[†,1] Thang Pham,[1] Anthony McFadden,[2] Nicholas Materise,[3] Paul Masih Das,[1] Matthew Cheng,[1] Dmitry Lebedev,[1] Stephanie M. Ribet,[1] Mitchell J. Walker,[1] David A. Garcia-Wetten,[1] Cameron J. Kopas,[4] Jayss Marshall,[4] Ella Lachman,[4] Nikolay Zhelev,[5,6,7] James A. Sauls,[8] Joshua Y. Mutus,[4] Corey Rae H. McRae,[2,9,10] Vinayak P. Dravid,[1,11] Michael J. Bedzyk,[1,6,*] Mark C. Hersam [1,12,13,*]

[1]Department of Materials Science and Engineering, Northwestern University, Evanston, IL 60208, USA

[2]National Institute of Standards and Technology, Boulder, CO 80305, USA

[3]Department of Physics, Colorado School of Mines, Golden, CO 80401, USA

[4]Rigetti Computing, Berkeley, CA 94710, USA

[5]Center for Applied Physics and Superconducting Technologies, Northwestern University, Evanston, IL 60208, USA

[6]Department of Physics and Astronomy, Northwestern University, Evanston, IL 60208, United USA

[7]Department of Physics, University of Oregon, Eugene, OR 97403

[8]Hearne Institute of Theoretical Physics, Department of Physics and Astronomy, Louisiana State University, Baton Rouge, LA 70803, USA

[9]Department of Physics, University of Colorado, Boulder, CO 80309, USA

[10]Department of Electrical, Computer, and Energy Engineering, University of Colorado, Boulder, CO 80309, USA

[11]Northwestern University Atomic and Nanoscale Characterization Experimental Center (NUANCE), Northwestern University, Evanston, IL 60208, USA

[12]Department of Chemistry, Northwestern University, Evanston, IL 60208, USA

[13]Department of Electrical and Computer Engineering, Northwestern University, Evanston, IL 60208, USA

†Contributed equally

*Corresponding authors: M.J.B. (bedzyk@northwestern.edu); M.C.H. (m-hersam@northwestern.edu)

## Abstract

Superconducting Nb thin films have recently attracted significant attention due to their utility for quantum information technologies. In the processing of Nb thin films, fluoride-based chemical etchants are commonly used to remove surface oxides that are known to affect superconducting quantum devices adversely. However, these same etchants can also introduce hydrogen to form Nb hydrides, potentially negatively impacting microwave loss performance. Here, we present comprehensive materials characterization of Nb hydrides formed in Nb thin films as a function of fluoride chemical treatments. In particular, secondary-ion mass spectrometry, X-ray scattering, and transmission electron microscopy reveal the spatial distribution and phase transformation of Nb hydrides. The rate of hydride formation is determined by the fluoride solution acidity and the etch rate of $Nb_2O_5$, which acts as a diffusion barrier for hydrogen into Nb. The resulting Nb hydrides are detrimental to Nb superconducting properties and lead to increased power-independent microwave loss in coplanar waveguide resonators. However, Nb hydrides do not correlate with two-level system loss or device aging mechanisms. Overall, this work provides insight into the formation of Nb hydrides and their role in microwave loss, thus guiding ongoing efforts to maximize coherence time in superconducting quantum devices.

## 1. Introduction

Superconducting qubits are of high interest for quantum information technologies due to their high gate fidelity and scalability.[1,2] Although the performance of superconducting qubits has increased dramatically over the past two decades, coherence times must increase further to achieve scalable quantum computing.[3-5] Among the possible sources of decoherence, fabrication processes are leading culprits since they can introduce impurities or defects within the qubits that host microwave loss mechanisms, such as two-level systems and excess unpaired quasiparticles.[1,6,7]

In an effort to leverage known fabrication processes and infrastructure, superconducting qubits are typically produced using clean room techniques that are borrowed from the complementary metal–oxide–semiconductor (CMOS) industry.[8,9] Niobium (Nb) has been used extensively as the primary superconductor in these qubits due to its compatibility with industrial-scale fabrication techniques and favorable superconducting properties, such as relatively high superconducting critical temperature ($T_c \sim 9.2$ K) and the corresponding low number of equilibrium quasiparticles at 10-20 mK.[10] However, in studies of both 2D resonators and 3D cavities, the native Nb surface oxide has been identified as a detrimental source of microwave loss.[11,12] Common fabrication protocols for surface cleaning and oxide removal employ wet chemical fluoride-based etchants, such as hydrofluoric acid (HF) and the associated buffered oxide etchant (BOE). These etchants have been adopted in multiple steps in qubit fabrication, including substrate cleaning before metallization and/or before Josephson junction (JJ) deposition, in addition to being employed just before cryogenic cooling in the case of resonator measurements.[9,13-15] In particular, it has been shown that using BOE to reduce amorphous surface oxides in Nb-based devices benefits resonator performance.[16] However, acidic solutions can also introduce hydrogen into bulk Nb once the $Nb_2O_5$ top layer is removed,[17] suggesting that Nb hydrides could result from fluoride-based etchants. The formation of Nb hydrides is potentially concerning for superconducting qubits since Nb hydrides have been identified as a critical contributor to reducing the quality factor in 3D superconducting radio frequency (SRF) Nb cavities, where Nb hydrides were introduced through a range of processing treatments to the Nb surface.[18-20] Recently, similar hydride features have also been identified in Nb 2D superconducting qubits.[21]

Here, we investigate the role of fluoride-based chemical etching processes in forming Nb hydrides in Nb thin films at room temperature and their subsequent effects on Nb coplanar waveguide (CPW) resonator performance. Various etchants are explored, including $NH_4F$, BOE (5:1, $NH_4F$:HF), and HF with a range of concentrations (2%, 5%, 8%, and 33%). Time-of-flight secondary ion mass spectrometry (ToF-SIMS) is then used to track the incorporation of hydride species into Nb thin films. In addition, the evolution of various crystallographic phases of Nb hydrides is characterized by X-ray diffraction (XRD). Transmission electron microscopy (TEM) techniques are further used to identify $NbH_x$ phases at nanoscale resolution. Through X-ray reflectivity (XRR) characterization, the etch rate of the amorphous native oxide $Nb_2O_5$ is tracked as a function of fluoride-based chemical etching, revealing that Nb hydride introduction is significantly accelerated following the full removal of $Nb_2O_5$. Moreover, the acidity of the etchant solution increases the hydrogen intake, as evidenced in the $NbH_x$ phase transitions. In terms of superconducting properties, the degree of Nb hydride formation is correlated with suppressed $T_c$ and residual resistance ratio (RRR). Additionally, microwave characterization of CPW resonators shows that the presence of Nb hydrides directly increases power-independent loss. On the other hand, two-level system (TLS) losses in the CPW resonators are not strongly affected by the presence of hydrides. Instead, the nominal changes in TLS loss are more likely associated with changes in $Nb_2O_5$ layer thickness caused by roughening of the Nb surface, as measured by atomic force microscopy (AFM). By elucidating the effects of fluoride-based etchants on hydride incorporation and superconducting properties, this work will help guide ongoing efforts to improve Nb superconducting quantum device performance.

## 2. Results and Discussion

To interrogate the incorporation of hydrogen into niobium from fluoride-based chemical etchants, we exposed ~80 nm thick Nb films deposited on Si (001) by DC sputtering (see details in Methods) to the following range of fluoride-based aqueous solutions listed in order of increasing acidity: NH4F, BOE (5:1, NH4F:HF), and HF (2%, 5%, 8%, 33%). Separate Nb thin-film samples were immersed in each etchant solution for 20 min and subsequently characterized after air exposure. Exposure to air minimizes further hydride formation after etching due to the formation of the native oxide $Nb_2O_5$, which acts as a hydrogen barrier.[17] ToF-SIMS was used to track the NbH⁻ ions as a function of depth into the film, as shown in **Figure 1A**. From the ToF-SIMS data, the NbH⁻ ion signal for the untreated control sample peaks immediately below the niobium surface oxide, composed primarily of $Nb_2O_5$, and rapidly drops off within the first ~ 8 nm of the film (**Figure S1**). This observation is consistent with a report by Lee *et al.*[21], in which the authors also found the presence of hydrides in Nb thin films with no etching processes. For the film treated with NH4F, ToF-SIMS shows characteristics almost identical to the control. Starting with the BOE sample, the NbH⁻ signal is detected deeper into the film, and this trend of increasing depth continues progressively as the etchant acidity increases. In the case of the untreated control, NH4F, BOE, and 2% HF samples, the hydride signal peaks and then rapidly decreases. On the other hand, for the 5% HF, 8% HF, and 33% HF samples, the hydride signal reaches a plateau, after which it decays at a slower rate. The extension of these plateaus is proportional to the hydrogen content introduced to the films. The decrease in intensity after the plateau could also indicate excess hydrogen escaping from the Nb film into the vacuum during the Cs⁺ etching required for SIMS measurements, which exposes bare Nb to vacuum after the native oxide removal.[22] Additionally, SIMS detected signals for other niobium hydride clusters, such as $NbH_2^-$ and $NbH_3^-$, follow the same trend as $NbH^-$ (**Figure S1**). Since all the etchants contain fluoride, we also examined the extent of $F^-$ ion incorporation into the films. However, the SIMS profiles for fluoride deviate strongly from the niobium hydride ion clusters, with only the most extreme case of 33% HF showing a significant degree of fluoride incorporation into the Nb thin film (**Figure S2**), although a weak increase is observed for the $F^-$ ion signal with increasing etchant solution acidity for the other cases.

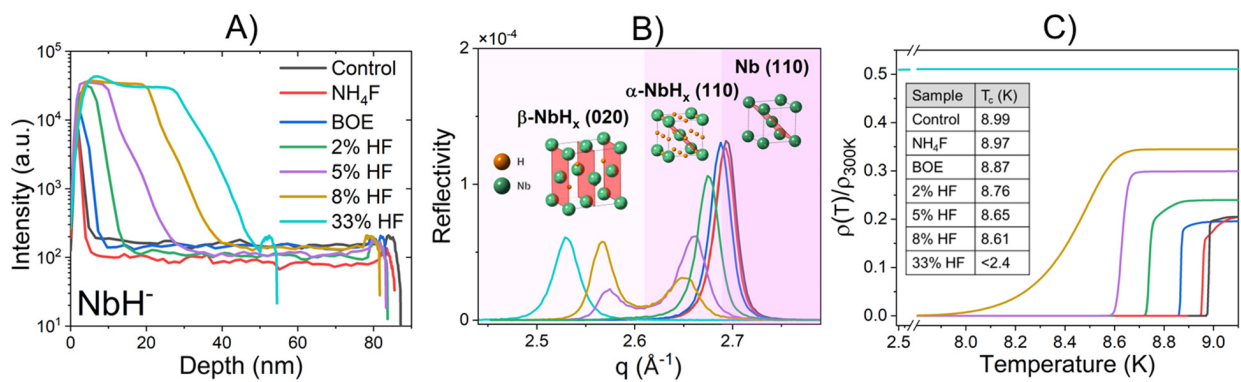

**Figure 1.** (A) NbH⁻ SIMS profiles, (B) XRD patterns, and (C) superconducting critical temperature measurements for a series of Nb thin films after 20 min fluoride-based etching treatments. The inset table in (C) shows the superconducting critical temperatures of the films. The line color legend in (A) also applies to (B) and (C).

The same series of samples were interrogated with XRD and XRR to examine further the effects of hydrogen incorporation on the Nb film structure (**Figure 1B and S3**). The XRD patterns show a shift in the Bragg peak towards lower scattering vector $q$ ($q=4\pi sin\theta/\lambda$) with increasingly aggressive etchant exposure, corresponding to an increase in the lattice spacing $d$ ($d=2\pi/q_{peak}$) of the Nb film. Using the change in the lattice spacing, the expansion of the lattice is tracked from Nb body-centered-cubic (BCC) metal to α-$NbH_x$ (BCC) as a result of interstitially dissolved hydrogen in tetrahedral sites. Furthermore, as hydrogen content is increased, a phase transition to β-$NbH_x$ face-centered-orthorhombic (FCO) is observed due to the ordering of the H atoms on a subset of tetrahedral interstitial sites.[23, 24] **Figure 1B** depicts the regions of hydride phases and suggests that hydrogen increasingly occupies tetrahedral sites of the BCC and FCO unit cells as the acidity of the etchant increases. Previously, the increase in lattice parameter in the Nb BCC structure as a function of hydrogen content was found to be 0.0023 Å per at.% of hydrogen.[25] **Table S1** depicts the volume fractions extracted from the Bragg peaks for Nb (110), α-$NbH_x$ (110), and β-$NbH_x$ (020) in addition to the approximated value of at.% of hydrogen based on the lattice expansion of α-$NbH_x$ (110).[26] In **Figure 1B**, the untreated and $NH_4F$-treated films have nearly identical Bragg peaks, suggesting minimal hydrogen incorporation from $NH_4F$ treatment. In the case of the BOE-treated sample, a slight expansion of the lattice is observed. The XRD peak shift is exacerbated in the 2% HF sample, indicating a significant α-$NbH_x$ component in the film. Beginning with the 5% HF treatment condition, substantial changes are observed in the film composition, as indicated by a secondary Bragg peak corresponding to β-$NbH_x$. This trend continues with the 8% HF treatment with the XRD pattern showing an increasing amount of β-$NbH_x$. In the case of 33% HF, the XRD pattern reveals that the film is entirely converted to β-$NbH_x$. In addition to XRD, XRR found an increase in $NbH_x$ concentration in the film with more aggressive etchants, as shown in the inset in **Figure S3**. In particular, the decreasing XRR critical angle corresponds to the electron density changing from Nb to NbH$_x$. This effect is evident in the 33% HF sample as XRR is sensitive to the β-NbH$_x$ phase, which corresponds to an electron density that is 90% of Nb. XRR also allows the thicknesses and interfacial roughnesses of the various layers in the film to be extracted including the surface oxide, bulk Nb, and Nb/Si substrate interface (**Table S2**). These values show that the thickness of the Nb films remains relatively constant for the different etchants, with a slight exception at 8% HF and more abruptly at 33% HF, where a considerable reduction in Nb film thickness is observed.

To interrogate the implications of hydride formation on superconducting properties, T$_c$ was extracted from temperature-dependent resistivity measurements as shown in **Figure 1C**. The trend in T$_c$ correlates well with the SIMS and XRD results, with the control and NH$_4$F samples being almost identical followed by a progressive depression in T$_c$ with increasing hydrogen incorporation. The depression of T$_c$ in the α-NbH$_x$ phase is explained by an increase in resistance due to interstitial hydrogen and a disruption of local superconductivity due to ε-NbH$_x$ precipitates formed at low temperatures.[27, 28] The 8% HF-treated film exhibited a significant deviation compared to the samples treated by less aggressive etchants, particularly a considerable broadening in the superconducting transition (ΔT) and increase in resistivity (ρ(T)/ρ$_{300K}$), due to the presence of an inhomogeneous system that consists of regions of superconducting phase (α-NbH) coupled with regions of non-superconducting phase (ε-NbH$_x$).[29] In the extreme limit of the 33% HF-treated film, a superconducting transition is not observed down to the measurement limit of 2.4 K, which is consistent with previous work that showed no sign of superconductivity for β-NbH$_x$/ε-NbH$_x$ down to 1.3 K.[28] The T$_c$, residual resistance ratio (RRR = R$_{300K}$/R$_{10K}$), and ΔT values are summarized in **Table S3.** The depression of T$_c$ and RRR, in addition to the broadening of ΔT, are all signatures of hydride formation compromising the superconducting properties of the Nb thin films.

Additional characterization was then performed to probe the time scales over which hydrogen incorporation occurs. For these experiments, thinner Nb films (~ 40 nm thick) were used to better detect changes in hydride concentration due to the smaller Nb overall volume. **Figure S4** shows the XRD and XRR overlays for all treatment conditions at 20 min for these thinner Nb. In a similar trend to **Figure 1**, the thinner Nb films also show a higher hydrogen loading as the etching solution acidity increases. From the XRD measurements for each etchant (**Figure S5**), the volume fraction of each phase was extracted from the Nb (110), α-NbH$_x$ (110), and β-NbH$_x$ (020) Bragg peaks at different etchant exposure times (45 s, 3 min, 20 min, 42 min, 2 h, and 13 h). **Figure 2** summarizes the evolution of these phases for the five different etchants. As before, the 33% HF condition fully converts the Nb film to β-NbH$_x$ after 20 min. For the 8% HF condition, after just 45 sec, evidence of α-NbH$_x$ (110) is already observed. However, after 20 min, a mix of hydride phases is still present, with complete conversion to β-NbH$_x$ occurring after 42 min. For the other less aggressive etchants, the onset of hydrogen incorporation and the phase transformation of the Nb film occurs over longer time scales. In particular, the 2% HF sample has a noticeable peak shift after 3 min, BOE after 20 min, and NH$_4$F only after significantly longer exposure (13 hr). In all cases, the formation of the β-NbH$_x$ phase proceeds after saturation of the α-NbH$_x$ phase. The decrease in the β-NbH$_x$ peak intensity for the samples treated with higher acidity etchants can be attributed to the etching of the Nb film and, thus, a reduction in the Nb film thickness. The etching rates extracted from XRR for the different etchants are provided in **Figure S6**.

The nanoscale distribution of NbH$_x$ phases in the 40-nm-thick Nb thin films is further elucidated using multimodal electron microscopy techniques (**Figure 3**). The analyzed samples in this case were the untreated control, 2% HF exposure for 20 min, BOE exposure for 42 min, and NH4F exposure for 13 hr, which has the highest hydrogen loading among these samples, according to the XRD results. We note that all Nb films have a columnar structure with grain sizes of 20-40 nm in width (**Figure S7A**). **Figure 3A** plots representative electron energy loss spectra (EELS) in the low-loss region of multiple samples after background subtraction (refer to **Figure S7B** for raw spectra before the background subtraction). All chemically treated samples show a shoulder at 5-8 eV (highlighted with a purple arrow), indicating metal-hydrogen bonding.[30-33] Conversely, the control sample has a distinct peak at ~10 eV, which is attributed to the Nb plasmon.[31] Additionally, the other Nb plasmon peak at ~21 eV in the control shifts toward higher energies (23-25 eV) in the processed samples, suggesting electron transfer between Nb and H atoms,[30, 31] which further confirms the presence of Nb-H bonding in the treated films. To investigate the nanoscale distribution of room-temperature $NbH_x$ phases detected by the XRD measurements, we employed 4D-STEM that acquires two-dimensional electron diffraction patterns at each probe scanning position,[34, 35] creating a spectrum image as shown in **Figure 3B**. This method collects structural information from a nanoscale volume, thus enabling the identification of $NbH_x$ phases embedded within the Nb matrix with sub-angstrom resolution. **Figures 3C** and **3E** show integrated 4D-STEM data (convergent-beam electron diffraction (CBED) patterns) summed over two areas in the 13 hr NH4F-treated sample enclosed by the blue and red boxes in **Figure 3B**, respectively. The experimental CBED patterns are compared with simulated patterns generated from Nb [110] and β-$NbH_x$ [010] (see Methods). The line profiles in **Figure 3D** comparing experimental and simulated Nb [110] patterns reveal a lattice expansion (~4.3 %) in the experimental line profile (see Methods), suggesting that the blue area in **Figure 3B** is an expanded Nb phase (i.e., α-$NbH_x$). On the other hand, the line profiles of the observed and simulated patterns for the red area in **Figure 3B** match reasonably well, indicating that the region is β-$NbH_x$ (**Figure 3F**). Other simulated CBED patterns of Nb and β-$NbH_x$ along different zone axes are provided in **Figure S8**. We note that some of the investigated regions were not transformed into $NbH_x$ (**Figure S8** shows examples of untransformed Nb regions). Additionally, the interface between the Si substrate and the Nb film is amorphous, and the film is polycrystalline with many grain boundaries, which together may help relieve mismatch strain between the substrate and the film in our samples. Consequently, the observed strain is attributed to the presence of α-NbHx.

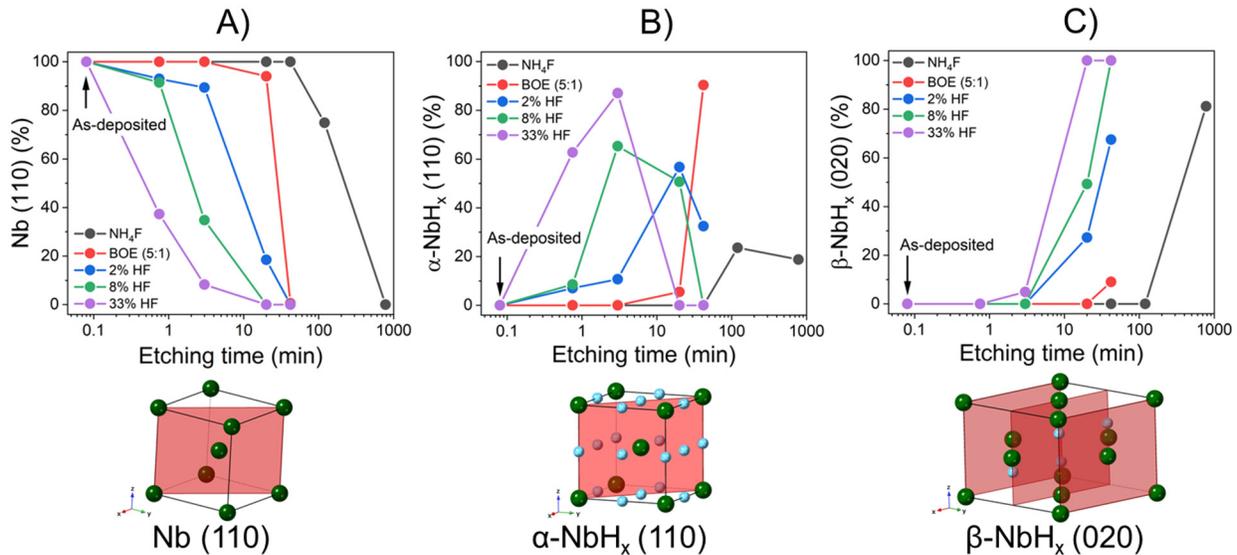

**Figure 2.** Volume fractions of (A) Nb (110), (B) α-NbH$_x$ (110), and (C) β-NbH$_x$ (020) of the 40-nm-thick Nb thin films treated with different etchants and times as extracted from XRD data of Figure S5. The volume fractions were calculated from the integrated area of the Bragg peaks using kinematical diffraction theory by multiplying the area by $V_{uc}/(LP*|F_H|^2)$, where $V_{uc}$ is the volume of the unit cell, LP is the Lorentz polarization factor and $F_H$ is the structure factor. The diffraction planes for each phase are shown below their respective phase volume fraction plots.

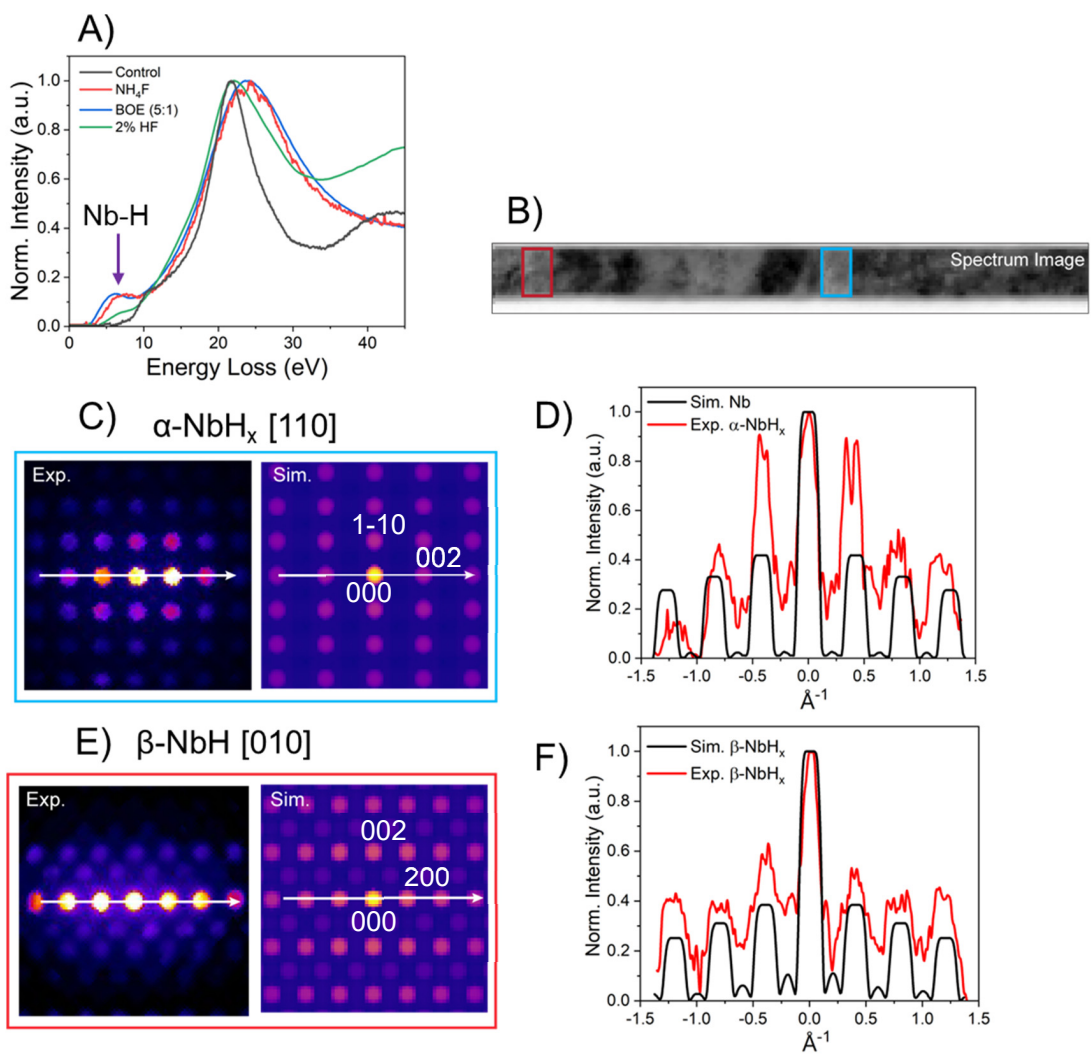

**Figure 3. Nanoscale distribution of NbHx phases.** (A) EELS of the untreated control and chemically treated Nb films. (B) A spectrum image of the NH$_4$F-treated Nb sample. The pixels within the blue and red boxes (20 × 35 nm$^2$) are used to generate the average CBED patterns in (C) and (E), respectively. (C) CBED patterns of Nb [110] were obtained by experiment (left) and multi-slice simulation (right). (D) Line profiles of patterns in (C) showing a lattice expansion in the experimental pattern. (E) CBED patterns of β-NbH [010] obtained by experiment (left) and multi-slice simulation (right). (F) Line profiles of patterns in (E).

To understand the role of the primary Nb surface oxide, Nb$_2$O$_5$, in incorporating hydrogen into the Nb films, stand-alone films of Nb$_2$O$_5$ were produced by pulsed laser deposition (PLD), which allowed the etch rate of the oxide to be probed for the various etchants used in this study. For the control sample, Nb$_2$O$_5$ is the most prominent oxide on the surface (**Figure S9A**). The PLD deposition was optimized using X-ray photoelectron spectroscopy (XPS) and XRR to track the film composition under different deposition conditions, as shown in **Figures S9B** and **S10**. To determine the etch rate of Nb$_2$O$_5$, the change in thickness of each film was measured with XRR during progressively longer etching treatments. The results shown in **Figure 4A** indicate that the etch rate of $Nb_2O_5$ increases with the acidity of the etchant and correlates strongly with the rate at which hydrogen is incorporated into the Nb film. To further investigate this behavior, a two-step sequential etch procedure was performed on a Nb film, namely a short etch (2 min) in 2% HF to remove the surface oxide and an immediate transfer to $NH_4F$ solution for more extended treatment (42 min). The XRD results (**Figure 4B**) show that while a film treated only with $NH_4F$ for 42 min remains unchanged, a film that is first etched in 2% HF and then in $NH_4F$ for 42 min now exhibits a mix of $\alpha$-$NbH_x$ and $\beta$-$NbH_x$. In particular, following a fast etch of $Nb_2O_5$ in 2% HF, the exposed Nb surface undergoes hydrogen incorporation upon exposure to $NH_4F$. Compared to the 42 min treatment using 2% HF only where mainly $\beta$-NbH is observed, less hydrogen is captured in the Nb for the $NH_4F$ case due to the lower concentration of $H^+$ in the $NH_4F$ solution. These findings were corroborated by XRR (**Figure 4C**), where a decrease in the electron density related to an increase in hydrogen loading was observed for the sequentially etched sample and, more severely, for the 42 min 2% HF sample. These results lead to the conclusion that the rate at which a particular etchant can incorporate hydrogen into a Nb film is dictated by both the rate of $Nb_2O_5$ removal and the available concentration of hydrogen in the etching solution.

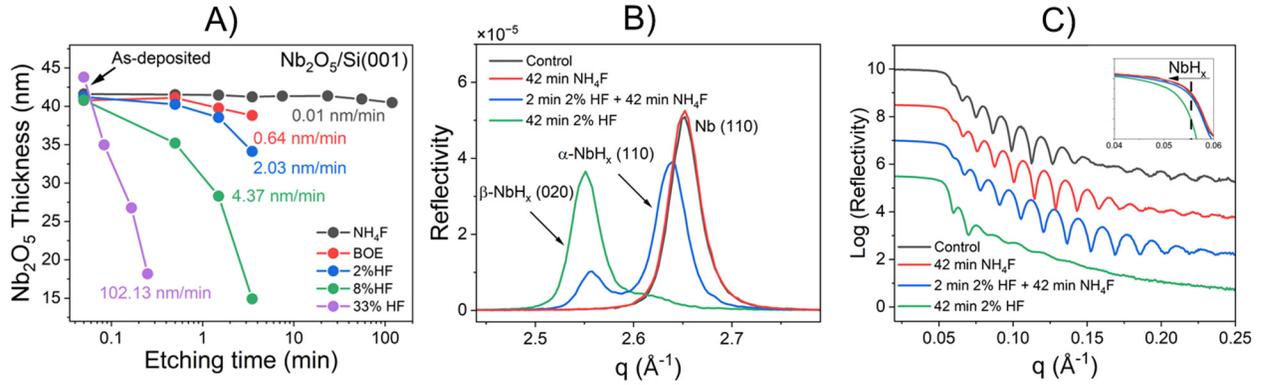

**Figure 4.** (A) XRR-determined etching rate of $Nb_2O_5$ thin films for different etchants. (B) XRD and (C) XRR of sequentially etched (2 min 2% HF and 42 min $NH_4F$) Nb thin films.

To evaluate the impact of Nb hydrides on superconducting devices, $\lambda/4$ CPW Nb resonators were fabricated. Superconducting microwave measurements of these resonators can effectively explore defects associated with losses.[6] The Nb resonators were treated with the same etching conditions as the aforementioned Nb films. Power-dependent measurements of the internal quality factor ($Q_i$) of these resonators were then measured after which power-independent ($\delta_{PI}$) and two-level system ($\delta_{TLS}$) loss tangents were extracted,[6] as shown in **Figure 5**. The microwave loss as a function of the number of photons (*S-curve*) is depicted in **Figure S11**. Five to eight resonators for each etching condition were measured. **Figure S12** shows the histograms for PI and TLS-associated losses. The PI loss data (**Figure 5A**) show a clear trend corresponding with increasing hydride concentration. In particular, the control, NH$_4$F, and BOE resonators all have similar degrees of PI losses, which agrees with the low hydrogen loading observed in these cases. However, starting with the 2% HF condition, an increase in PI loss was observed that continues to increase with the 5% HF and 8% HF samples, where the latter case shows an increase in the PI loss tangent by over a factor of four compared to the control. We also tested 33% HF-treated resonators but could not measure a signal from those samples, which suggests that even at the dilution cryostat base temperature (measurements performed at 13 mK), the 33% HF-treated Nb is either still not superconducting or has an internal quality factor so low that resonators could not be measured. Similarly, our temperature-dependent resistivity measurements (**Figure 1C**) for the equivalent 33% HF-treated Nb thin-film sample showed no sign of a superconducting transition down to 2.4 K.

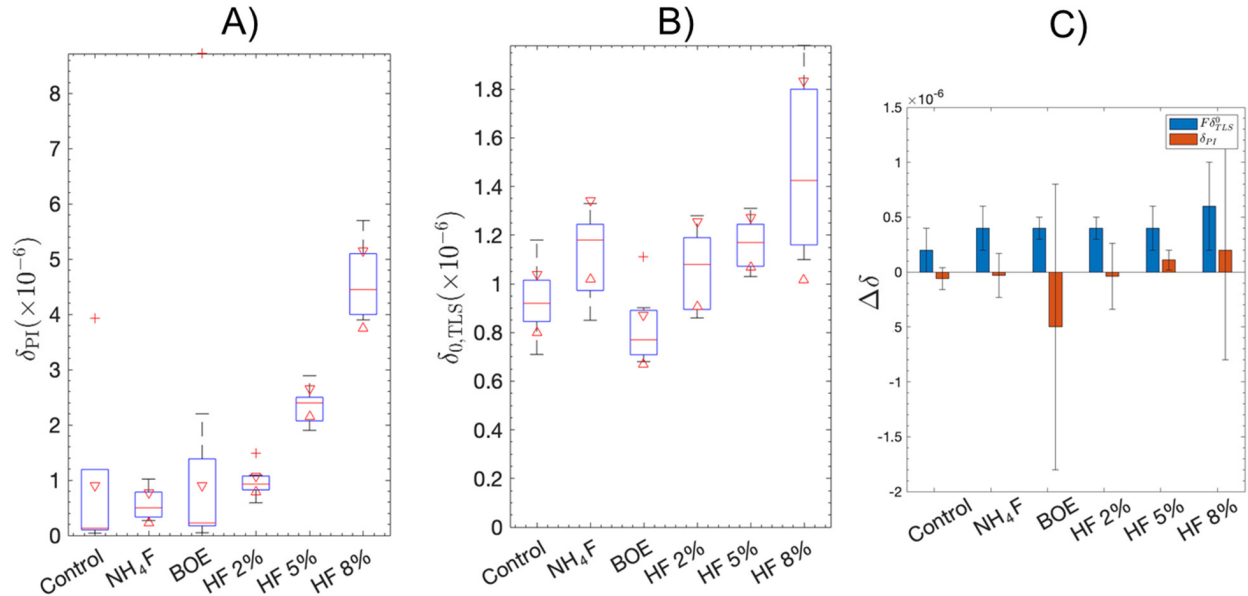

**Figure 5.** (A) Power-independent (PI) and (B) two-level systems (TLS) losses of Nb CPW resonators after 20 min treatments in the indicated etchants. Losses were extracted from the nonlinear least square fit to the TLS power-dependent model. The blue boxes contain the 25$^{th}$ and 75$^{th}$ percentiles, the red horizontal lines indicate the median values, the red triangles show the 95% confidence intervals about the median, the red plus signs are outliers, and the black horizontal lines are the minimum and maximum values excluding outliers. (C) Change in PI and TLS losses after two months of aging in air.

The increased PI loss along with the $T_c$ depressions discussed above indicates increased quasiparticle generation due to higher hydrogen loading. Several possible mechanisms could explain this increased quasiparticle generation. One possibility is that the hydrides are acting as magnetic scattering impurities, although the magnetism of $NbH_x$ remains an open question in the field. In addition, the increase in quasiparticles and the reduction in $T_c$ as the presence of $NbH_x$ increases are consistent with recent theoretical results by Sauls and collaborators.[36] Their study indicates that due to the anisotropic superfluid gap of Nb, impurities (e.g., $NbH_x$ precipitates) can still act as pair-breakers even when non-magnetic. Lastly, another theoretical result by Sauls and collaborators shows that TLSs embedded in bulk superconductors can generate dissipative quasiparticles at GHz frequencies.[37] Since the $T_c$ and resonator data suggest that the β-$NbH_x$ phase is not superconducting, increased quasiparticle generation could also result from these normal metal components of the film.

The impact of Nb hydrides on TLS losses is less pronounced (**Figure 5B**). In the case of the control, $NH_4F$, 2% HF, and 5% HF samples, similar levels of TLS loss are observed. With the BOE treatment condition, a slight shift downward is apparent in the TLS losses, likely due to the BOE etchant reducing the thickness of the Nb surface oxide while introducing minimal hydrides. For the 8% HF treatment condition, an increase in TLS and an increase in the spread of TLS loss tangents are observed. These effects are consistent with a rise in the surface roughness of the Nb film and an associated increase in the surface oxide.

AFM was used to characterize the surface of the Nb and Si areas on the fabricated resonator chips after each treatment protocol (**Figure S13**). From these AFM images, the RMS surface roughness and the surface area divided by the projected area (constant= 4 $μm^2$) were extracted (**Figure S14**). Previously, a decrease in $Nb_2O_5$ thickness has been linked to an overall improvement in resonator loss.[16] However, in that study, both $Nb_2O_5$ and $SiO_2$ were found to host TLS losses. We expect the $SiO_2$ in our samples to be uniform across the different resonators due to the similarity in Si surface roughness (**Figure S14**) and since the samples were exposed to air for the same time before microwave measurements. In the case of the Si surface, a slight smoothening effect is observed with increasing etchant acidity, but the overall change in the surface area is minimal. In contrast, with the Nb surface, a substantial roughening effect is observed with increasing etchant acidity, which results in more $Nb_2O_5$ that is expected to increase

TLS loss. The increase in the oxide thickness was further corroborated by the XRR fittings (**Table S3**) for the 8% HF and 33% HF samples, as the oxide is formed on a rougher Nb surface caused by the etching of the film.

Lastly, a second series of resonator measurements was performed to determine if any aging effects could be linked to hydrogen loading. The set of Nb resonators measured in **Figures 5A** and **5B** was aged over two months in ambient laboratory conditions, and then microwave characterization was repeated. **Figure 5C** shows the difference between PI and TLS losses before and after aging. **Figure S15** provides an overlay of PI and TLS loss tangents for both measurement series. The PI losses change minimally after aging, suggesting no additional aging effects from hydrides related to PI loss. This result is consistent with the hydrogen loading remaining unaltered throughout the aging process due to the native oxide $Nb_2O_5$ preventing both further hydrogen intake and hydrogen escaping from the film.[17] For TLS losses, a uniform increase was observed in all resonators for all treatment conditions following aging. These data are consistent with increased TLS loss sources due to oxide thickening[11] and additional contamination during storage.[38] Previous reports have also observed increased TLS losses after long exposure of Nb to air.[11] It should be noted that earlier studies on Nb cavities and 2D Nb qubits have shown that hydride precipitates disorder the Nb surface,[19, 21] which has led to hydrides being hypothesized to be sources for both TLS and PI losses in addition to contributing to aging mechanisms.[21] However, our results suggest that while the amount of Nb hydride correlates with increased PI losses, it does not contribute significantly to TLS losses or aging.

## 3. Conclusions

In this study, we investigated the role of fluorine-based etchants in forming Nb hydrides in Nb thin films. For this purpose, we employed a range of etching solutions: $NH_4F$, BOE, and HF (2%, 5%, 8%, and 33%). As the etchant solution acidity increases, more hydrides are detected using SIMS. Additionally, phase transitions from niobium metal to α-$NbH_x$ and ultimately to β-$NbH_x$ are observed by X-ray and electron diffraction with increasingly aggressive etching conditions. Moreover, we extracted power-dependent loss tangents from measurements of Nb CPW resonators with a range of hydrogen loading. For power-independent loss, a strong correlation was observed with the hydrogen loading level. On the other hand, TLS loss is essentially unaffected by hydrogen loading, although the roughening of the Nb surface by aggressive etching does cause an increase in the amount of surface $Nb_2O_5$ and associated TLS losses. The aging of these resonators in air showed a stable PI loss as the hydride concentration remained constant and a uniform increase in TLS losses for all hydrogen loading conditions due to additional oxide growth and adventitious contamination during storage.

Additionally, based on our study of $Nb_2O_5$ films, we determined that the rate of hydride formation is directly correlated to the etch rate of $Nb_2O_5$ for a given etching solution. Combining this finding with our results on the effect of sequential etching with 2% HF and $NH_4F$, two main factors contribute to hydrogen incorporation into Nb films: (1) Removal of the $Nb_2O_5$ surface oxide, which acts as a diffusion barrier to hydrogen; (2) Concentration of hydrogen in the etching solution to which the oxide-free Nb surface is exposed. Therefore, cleaning and etching strategies should consider these two factors to minimize the formation of detrimental hydrides in superconducting devices. In this manner, this comprehensive study on the formation of Nb hydrides and their correlation to loss mechanisms sheds light on the relationship between materials processing and superconducting device performance, thus guiding the development of fabrication procedures for quantum information technologies.

## 4. Methods

**Thin film preparation:** Nb thin films for structural and chemical characterization and further cryogenic microwave measurements (80 nm thick) were deposited by DC magnetron sputtering. First, double-side polished intrinsic Si (001) wafers (WaferPro 380 μm, >10 MΩ) were prepared using standard RCA1/RCA2 cleaning. Before Nb deposition, the wafers were immersed in 6:1 buffered oxide etch for 2 min, rinsed in deionized (DI) water, and dried in $N_2$ gas. Next, niobium was deposited in a Multi Tool Deposition System (PVD Products) chamber with base pressure $<1\times10^{-8}$ Torr. The deposition was performed at room temperature with 3 mTorr of Ar and 300 W from a 2-inch diameter Nb target (JX Nippon, 5N), resulting in a 1.7 Å/s deposition rate.

Nb films for interfacial/bulk studies (40 nm thick) were deposited by high-power impulse magnetron sputtering (HiPIMS) on intrinsic 6-inch Si (001) wafers. The substrates were sequentially cleaned using RCA1, buffered oxide etchant (5:1), and RCA2. Before deposition, an additional BOE (5:1) treatment for 30 sec was employed to remove the substrate surface oxide. A system with base pressure ~2x10$^{-9}$ Torr was used to sputter a 5N pure Nb target. The substrate received a bakeout at 150 °C in the loadlock before transfer into the main deposition chamber. These samples present an initial out-of-plain strain of 1.3%, which is calculated from the experimental Nb (110) Bragg peak in the XRD scan and compared to Nb bulk values.

The niobium pentoxide ($Nb_2O_5$) thin films were deposited by pulsed laser deposition (PLD) on an intrinsic Si (001) wafer. PLD was performed in a PVD Products PLD/MBE 2300 equipped with a 248 nm KrF excimer laser. The films were deposited from a $Nb_2O_5$ target at room temperature in a 10 mTorr $O_2$ atmosphere and a pulse frequency of 10 Hz. The laser was focused on a 2 × 4 mm$^2$ spot size with an energy of ~160 mJ/pulse.

The etching treatments were performed using different HF (KMG Electronic Chemicals, 49% aqueous) and $NH_4F$ (KMG Chemicals, 40% Aqueous) solutions at room temperature: $NH_4F$, BOE (5:1), 2% HF, 5% HF, 8% HF, and 33% HF in descending pH order. The 80-nm-thick Nb films were etched for 20 min in all conditions at room temperature. The 40-nm-thick Nb and $Nb_2O_5$ films were etched for different times and specific etchants to measure structural and interfacial changes. All samples were etched at room temperature on an orbital shaker rotating at 60 rpm to maintain uniform etching. All thin film samples were approximately 10 × 10 mm$^2$ in area. TEM characterization was performed on the 40-nm-thick Nb samples etched with the following conditions: (1) control; (2) 42 min BOE; (3) 25 min 2% HF; (4) 13 hr $NH_4F$.

**X-ray diffraction and reflectivity:** Specular X-ray reflectivity (XRR) and X-ray diffraction (XRD) were performed with a Smartlab Gen 2 diffractometer equipped with a 9 kW Cu rotating anode and a Ge (220) 2-bounce monochromator ($\lambda$ = 1.5406 Å). The XRR data were corrected for beam footprint and non-specular background signal. Motofit software was used to fit the XRR data with a multiple-slab model. All the XRR and XRD data were normalized with the straight-to-beam intensity. The Nb, α-NbH$_x$, and β-NbH$_x$ phases were determined by fitting Gaussian curves in the three 2θ regions of interest. The phase volume was determined from kinematical theory, integrating the areas from the Gaussian fits and considering each phase's differential cross-section, including the unit cell structure, unit cell volume, and polarization factor.[26] Specifically, the Bragg peak area was multiplied by $V_{uc}/(LP*|FF|^2)$, where $V_{uc}$ is the volume of the unit cell, LP is the Lorentz polarization factor, and FF is the atomic form factor.

**Transmission electron microscopy:** Lamella cross-sections were prepared using a dual-beam focused ion beam (FIB) FEI-SEM Helios Nanolab. The lift-off was performed using 30 kV Ga$^+$ ions, and the final cleaning step was carried out at 2 kV to remove surface damage on the area of interest, resulting in a final sample thickness of 50-100 nm. Transmission electron microscopy (TEM) and scanning transmission electron microscopy (STEM) data were collected on a probe-corrected JEOL ARM200 S/TEM operating at 200 kV. The convergent angle for ADF-STEM imaging was 25 mrad. Electron energy loss spectra (EELS) were acquired at 200 kV using a Gatan GIF Quantum on a K2 pixelated detector. Data processing (background subtraction, plural scattering removal, and signal mapping) was conducted using Gatan GMS software. Four-dimensional STEM (4D-STEM) data were collected by a OneView camera with a 512 × 512 or 256 × 256 pixels software-bin size. The convergent angles were 3-5 mrad, and the camera length was 20 cm. The step size was 5-10 Å. 4D-STEM patterns were simulated using the abTEM package with input parameters from the experimental conditions. As shown in **Figure 3D and Figure 3F**, we plot the line profiles showing intensities of diffracted spots as the function of its reciprocal distance from the center (000) spot. The lattice parameters ($d$) can be calculated by $d_{hkl}=1/g_{hkl}$ where $g_{hkl}$ is the length from (000) to the (hkl) spot. The lattice parameters corresponding to (001) planes and the lattice expansion are calculated for both experimental α-NbH$_x$ and simulated Nb as follows:

$g_{\alpha\text{-NbHx}} = 0.5814$ Å-1 → $d_{\alpha\text{-NbHx}} = 1/0.5814 = 1.7199$ Å

$g_{sim\text{-Nb}} = 0.6066$ Å-1 → $d_{sim\text{-Nb}} = 1/0.6066 = 1.6485$ Å

The lattice expansion for the (001) planes is therefore: $d_{\alpha\text{-NbHx}} - d_{sim\text{-Nb}} / d_{sim\text{-Nb}} = 4.3\%$

**Atomic force microscopy:** An Asylum Cypher atomic force microscope with a Si cantilever (resonant frequency 320-340 kHz) was used in tapping mode. The image resolution was 512 × 512 pixels at a scanning rate of 1 Hz.

**X-ray photoelectron spectroscopy:** Spectra were collected in a Thermo Scientific ESCALAB 250Xi XPS spectrometer equipped with a monochromated Al Kα X-ray source with an energy of 1486.6 eV. The measurement spot size was ~500 μm, and a flood gun for charge compensation was utilized. The analysis was performed using Avantage (Thermo Scientific) software. The core levels were fitted using a modified Shirley background and a Gaussian-Lorentzian product (70%

Gaussian and 30% Lorentzian). All the peaks were charged-corrected to adventitious carbon (C 1s) at 284.8 eV.

**Time-of-flight secondary-ion mass spectrometry:** An IONTOF M6 dual-beam system (IONTOF GmbH) was used to analyze the concentrations and depth distribution of the Nb thin films. The measurements were performed in negative polarity using $Bi^+$ ions at 30 keV on a 25 × 25 $\mu m^2$ area. In addition, $Cs^+$ ions with an energy of 500 eV were used to sputter 150 × 150 $\mu m^2$ area for depth profiling. SurfaceLab7 software was used to plot and analyze the SIMS data.

**Superconducting transition temperature characterization:** Electrical transport measurements were performed using a Quantum Design PPMS (Dynacool) in a four-probe geometry. The devices were wire-bonded to the chip carriers using a homebuilt In-Au bonder. The critical temperature ($T_c$) was calculated from the intersection of two linear fits at and after the sharp transition. The residual resistance ratio (RRR) was calculated from the ratio of the resistances at 300 K and 10 K. The transition broadening ($\Delta T$) was calculated from the temperature interval between the two points associated with 10% and 90% of the resistance taken at the superconducting phase transition.

**Niobium resonator fabrication:** The 80 nm Nb films were patterned into coplanar waveguide (CPW) resonators using standard photolithography and dry etching. The Nb-coated wafer was first solvent cleaned to obtain a consistent starting surface, followed by a spin coating of P20 primer and SPR660 photoresist, and finally soft baking at 95 °C for 1 min. The wafer was then patterned using direct laser writing, baked post-exposure at 110 °C for 1 min, and developed in MF26A. An established CPW resonator design was utilized that consisted of eight frequency-multiplexed, inductively-coupled, quarter-wave resonators in hanger mode off a central feedline.[39] The CPW conductor/gap dimensions were 6 μm/3 μm such that the resonance frequencies fell between 4-8 GHz. Following photolithography, the wafer was cleaned in an oxygen plasma and etched in an inductively coupled plasma (ICP) tool using $SF_6$. The surface was monitored optically and timed to etch through the Nb and approximately 100 nm into the Si substrate. After etching, the photoresist was removed using an NMP-based remover held at 80 °C with ultrasonication followed by rinsing in isopropanol. The finished wafer was then coated in photoresist to protect the surface during dicing. The Nb resonators were treated for 20 min in different solutions before measurements.

**Resonator microwave characterization:** We performed microwave transmission measurements of coplanar waveguide hanger resonators using a vector network analyzer over several decades of input power. The resonators were packaged in gold-plated, oxygen-free copper sample boxes with pogo pins to hold the samples in place and connect the ground planes to the boxes. The samples were first measured in a single cooldown in a cryogen-free dilution refrigerator with a base temperature of 13 mK. Later, the resonators were measured again in a second cooldown after two months of exposure to air. The transmission data were fitted using the diameter correction method (DCM)[40] with a circle fit normalization routine[41] to account for asymmetric resonances, thus allowing the internal quality factors to be extracted as a function of input power. A secondary fit of inverse quality factors versus the estimated number of photons[42] in each resonator to a two-level system (TLS) model provided the power-independent and TLS losses.[43] The formula used to fit the power-dependent loss model is as follows:

$$\delta = F\delta_{TLS}^0 \frac{\tanh \frac{\hbar\omega}{2k_B T}}{\left[1 + \left(\frac{n}{n_c}\right)\right]^\beta} + \delta_{PI}$$

where $\delta$ is the total loss, $F$ is the filling factor, $\delta_{TLS}^0$ is the intrinsic TLS loss, $\omega$ is the frequency, $k_B$ is the Boltzmann constant, $\langle n \rangle$ is the average number of photons at the resonator, $n_c$ is the critical photon number of the TLS population, $\beta$ is the phenomenological scaling factor around 0.5, and $\delta_{PI}$ is the power-independent loss.

**Statistical Analysis:** Resonator transmission data was preprocessed using a circle fit normalization and then analyzed using a Lorentzian fit.[41] Losses were extracted from the nonlinear least square fit to the TLS power-dependent model. The software to perform the measurements, DCM fit, and TLS fit is available on GitHub: https://github.com/Boulder-Cryogenic-Quantum-Testbed/scresonators. Five to eight resonators for each etching condition were measured in the two cool-downs. The resonator data is presented in 25th and 75th percentiles, median, outliers, and maximum and minimum values, excluding such outliers.

# Acknowledgments

This work was primarily supported by the U.S. Department of Energy, Office of Science, National Quantum Information Science Research Centers, Superconducting Quantum Materials and Systems Center (SQMS) under contract No. DE-AC02-07CH11359. This work made use of the Jerome B. Cohen X-Ray Diffraction Facility supported by the MRSEC program of the National Science Foundation (DMR-2308691) at the Materials Research Center of Northwestern University and the Soft and Hybrid Nanotechnology Experimental (SHyNE) Resource (NSF ECCS-1542205.) This work also made use of the EPIC and Keck-II facilities of the Northwestern University NUANCE Center, which has received support from the SHyNE Resource (NSF ECCS-2025633), the IIN, and the Northwestern MRSEC program (NSF DMR-2308691). S.M.R. also acknowledges support from the International Institute of Nanotechnology and 3M.

# Supplementary Information

## Formation and Microwave Losses of Hydrides in Superconducting Niobium Thin Films Resulting from Fluoride Chemical Processing

Carlos G. Torres-Castanedo,[†,1] Dominic P. Goronzy,[†,1] Thang Pham,[1] Anthony McFadden,[2] Nicholas Materise,[3] Paul Masih Das,[1] Matthew Cheng,[1] Dmitry Lebedev,[1] Stephanie M. Ribet,[1] Mitchell J. Walker,[1] David A. Garcia-Wetten,[1] Cameron J. Kopas,[4] Jayss Marshall,[4] Ella Lachman,[4] Nikolay Zhelev,[5,6,7] James A. Sauls,[8] Joshua Y. Mutus,[4] Corey Rae H. McRae,[2,9,10] Vinayak P. Dravid,[1,11] Michael J. Bedzyk,[1,6,*] Mark C. Hersam [1,12,13,*]

[1]*Department of Materials Science and Engineering, Northwestern University, Evanston, IL 60208, USA*

[2]*National Institute of Standards and Technology, Boulder, CO 80305, USA*

[3]*Department of Physics, Colorado School of Mines, Golden, CO 80401, USA*

[4]*Rigetti Computing, Berkeley, CA 94710, USA*

[5]*Center for Applied Physics and Superconducting Technologies, Northwestern University, Evanston, IL 60208, USA*

[6]*Department of Physics and Astronomy, Northwestern University, Evanston, IL 60208, United USA*

[7]*Department of Physics, University of Oregon, Eugene, OR 97403*

[8]*Hearne Institute of Theoretical Physics, Department of Physics and Astronomy, Louisiana State University, Baton Rouge, LA 70803, USA*

[9]*Department of Physics, University of Colorado, Boulder, CO 80309, USA*

[10]*Department of Electrical, Computer, and Energy Engineering, University of Colorado, Boulder, CO 80309, USA*

[11]*Northwestern University Atomic and Nanoscale Characterization Experimental Center (NUANCE), Northwestern University, Evanston, IL 60208, USA*

[12]*Department of Chemistry, Northwestern University, Evanston, IL 60208, USA*

[13]*Department of Electrical and Computer Engineering, Northwestern University, Evanston, IL 60208, USA*

*†Contributed equally*

*\*Corresponding authors: M.J.B. ([bedzyk@northwestern.edu](bedzyk@northwestern.edu)); M.C.H. ([m-hersam@northwestern.edu](m-hersam@northwestern.edu))*

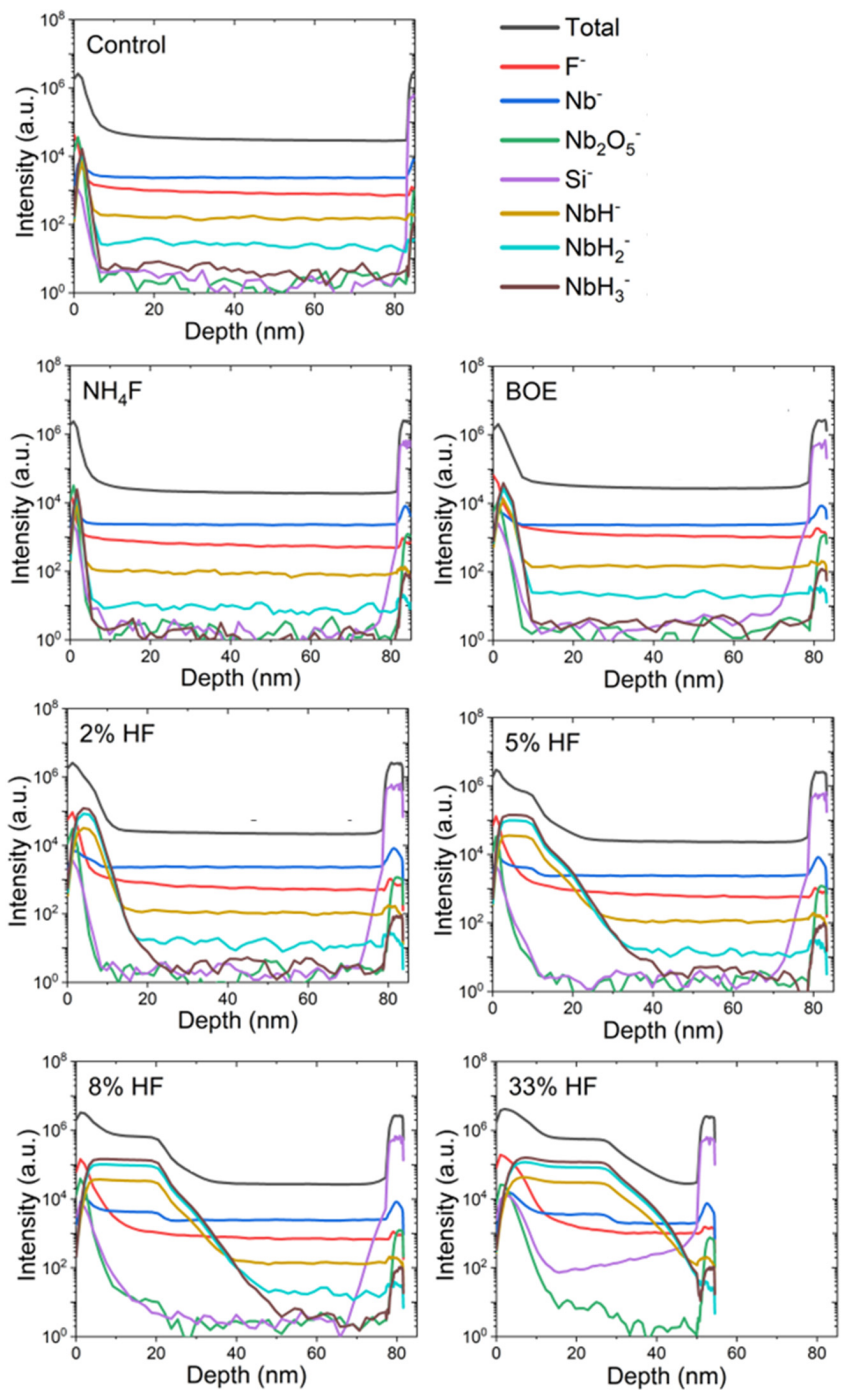

**Figure S1.** Secondary ion mass spectrometry (SIMS) profiles of Nb thin films after 20 min etch treatments for selected ions.

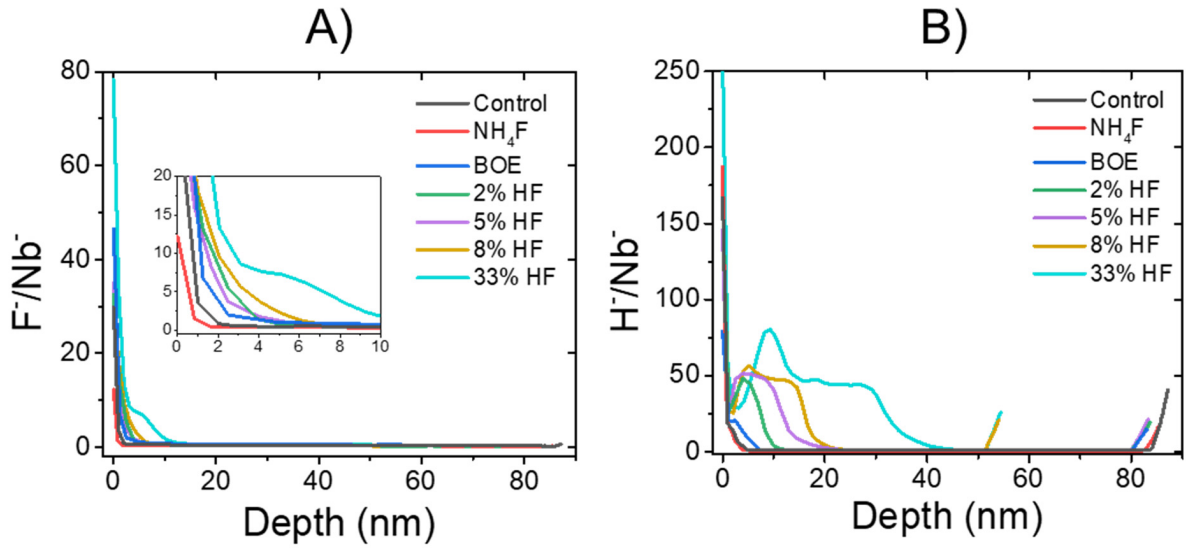

**Figure S2** (a) $F^-/Nb^-$ and (b) $H^-/Nb^-$ SIMS profiles of Nb thin films after 20 min wet etch treatments.

| Sample | Nb (110) | α-NbH$_x$(110) x=0.005-0.25 | β-NbH$_x$ (020) X>0.80 |
|---|---|---|---|
| Control | 100% | | |
| NH$_4$F | 100% | | |
| BOE (5:1) | 86.4% | 13.6% (x≈0.03) | |
| 2% HF | | 100% (x≈0.10) | |
| 5% HF | | 84.7% (x≈0.19) | 15.3% |
| 8% HF | | 62.7% (x≈0.25) | 37.3% |
| 33% HF | | | 100% |

**Table S1.** Volume fractions for the Nb and NbH$_x$ phases of the Nb thin films etched for 20 min. Fractions extracted from the integrated area of the Bragg peaks and using kinematical diffraction theory. The volume fraction was calculated by multiplying the area by $V_{uc}/(LP*|FF|^2)$, where $V_{uc}$ is the volume of the unit cell, LP is the Lorentz polarization factor, and FF is the atomic form factor.

| Sample | Nb$_x$Si$_y$ | Nb | Nb$_x$O$_y$ | Total |
|---|---|---|---|---|
| Control | 1.3 | 79.8 | 3.0 | 84.1 |
| NH$_4$F | 1.3 | 78.9 | 2.5 | 82.7 |
| BOE (5:1) | 1.3 | 76.2 | 2.5 | 80.0 |
| 2% HF | 1.3 | 76.0 | 2.5 | 79.8 |
| 5% HF | 1.3 | 76.1 | 2.5 | 79.9 |
| 8% HF | 1.3 | 72.9 | 4.3 | 78.5 |
| 33% HF | 1.3 | 43.8 | 6.2 | 51.3 |

**Table S2.** Nanometer (nm) thicknesses extracted from X-ray reflectivity (XRR) fittings of the Nb thin films etched for 20 min.

| Sample | T$_c$ (K) | RRR | ΔT (K) |
|---|---|---|---|
| Control | 8.99 | 5.08 | 0.015 |
| NH$_4$F | 8.97 | 5.64 | 0.015 |
| BOE | 8.87 | 5.13 | 0.033 |
| 2% HF | 8.76 | 4.18 | 0.035 |
| 5% HF | 8.65 | 3.34 | 0.065 |
| 8% HF | 8.61 | 3.09 | 0.465 |
| 33% HF | <2.4 | 1.96 | - |

**Table S3.** Critical temperature, residual resistance ratio (RRR=R$_{300K}$/R$_{10K}$), and transition width of Nb thin films etched for 20 min.

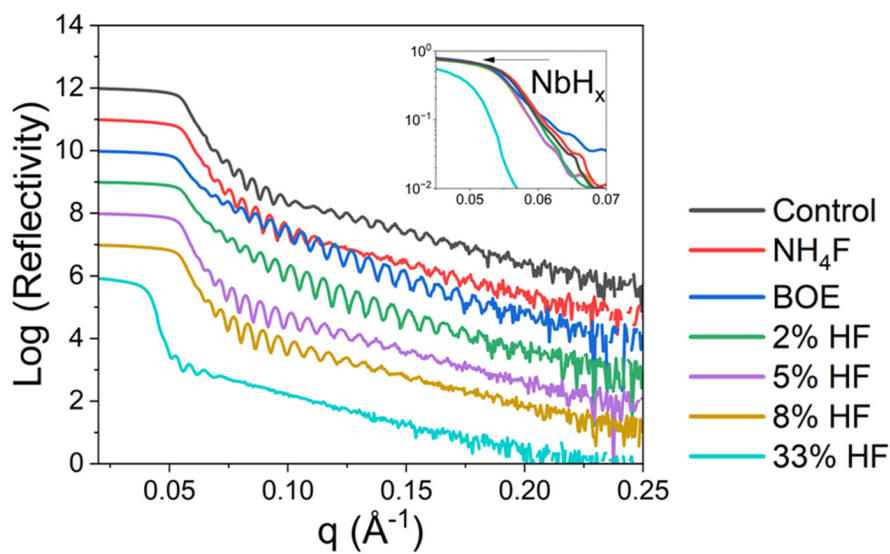

**Figure S3.** XRR of Nb thin films after 20 min etching treatments. The inset shows the decrease of the critical angle from Nb to NbH$_x$. Each profile has a constant offset.

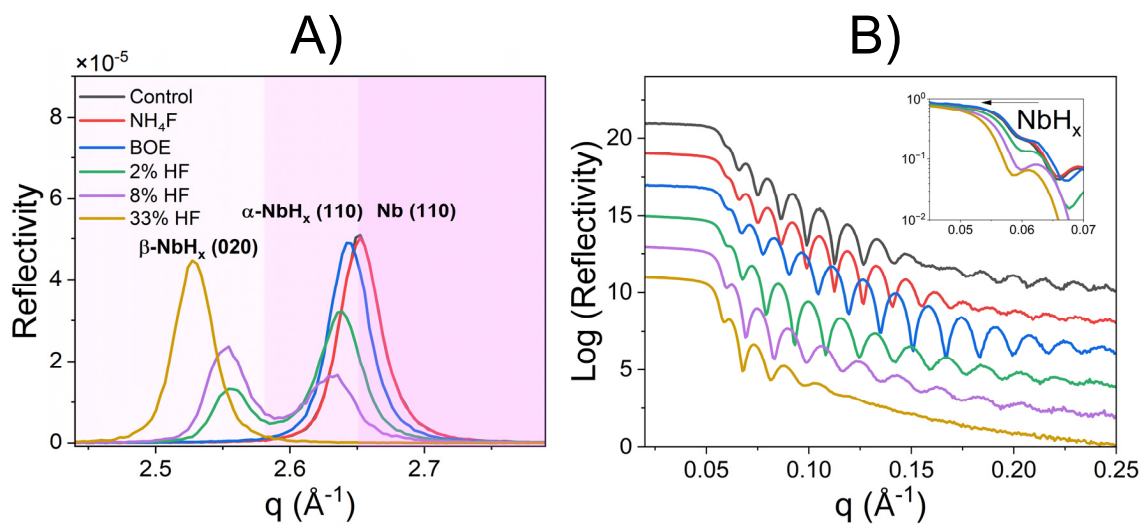

**Figure S4.** (A) X-ray diffraction (XRD) and (B) X-ray reflectivity (XRR) of 40-nm-thick Nb thin films after 20 min etching treatments. The inset shows the decrease of the critical angle from Nb to NbH$_x$.

S5

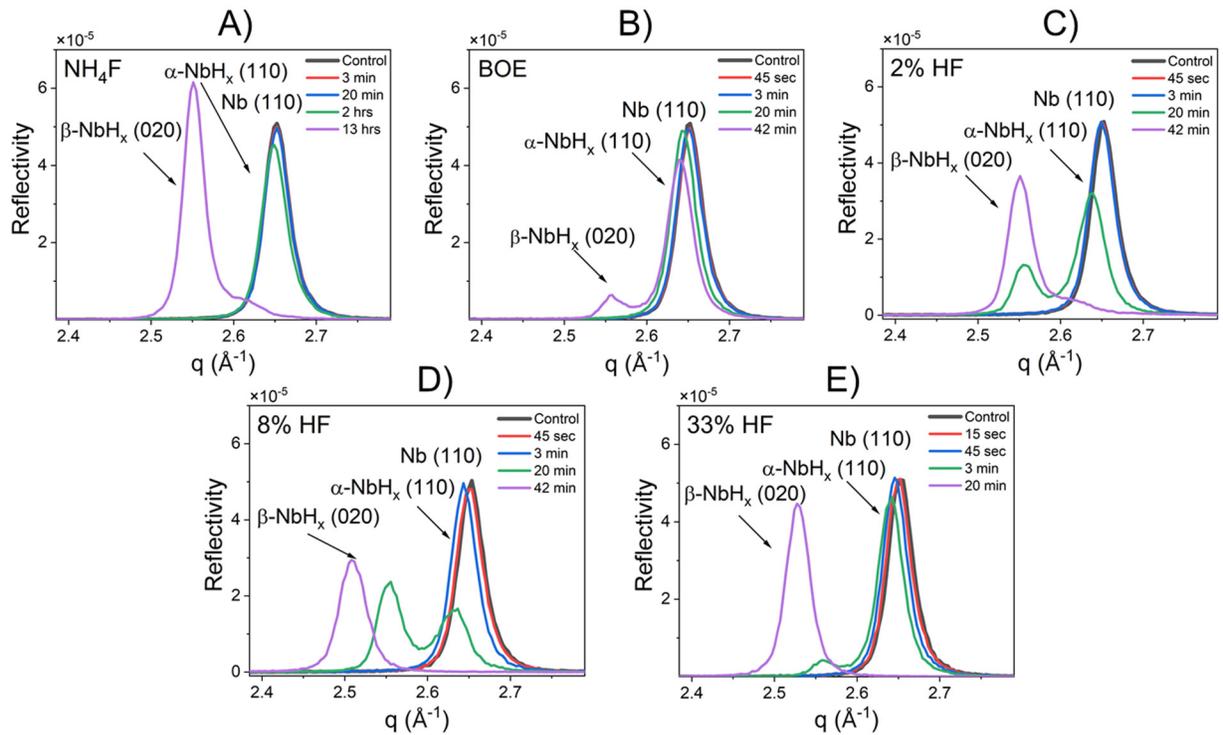

**Figure S5.** XRD of 40 nm Nb thin films treated with different etching times and etchants including: (A) NH$_4$F; (B) BOE; (C) 2% HF; (D) 8% HF; (E) 33% HF.

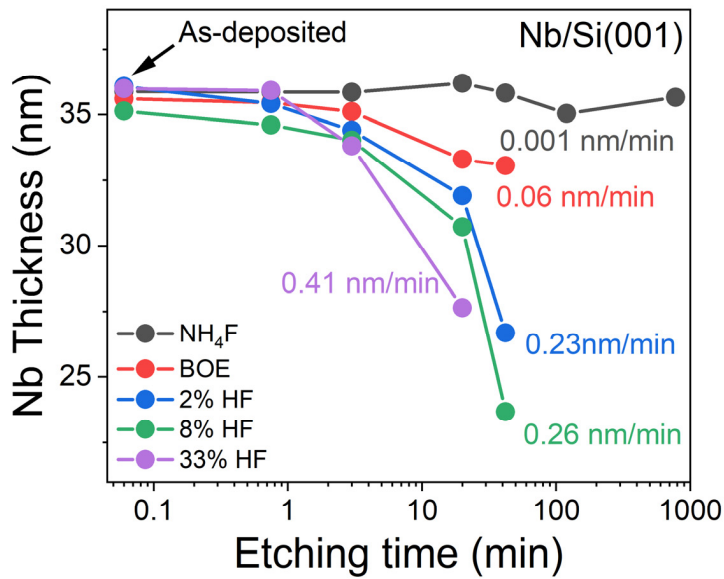

**Figure S6.** Etching rate of Nb thin films for different etchants extracted from XRR.

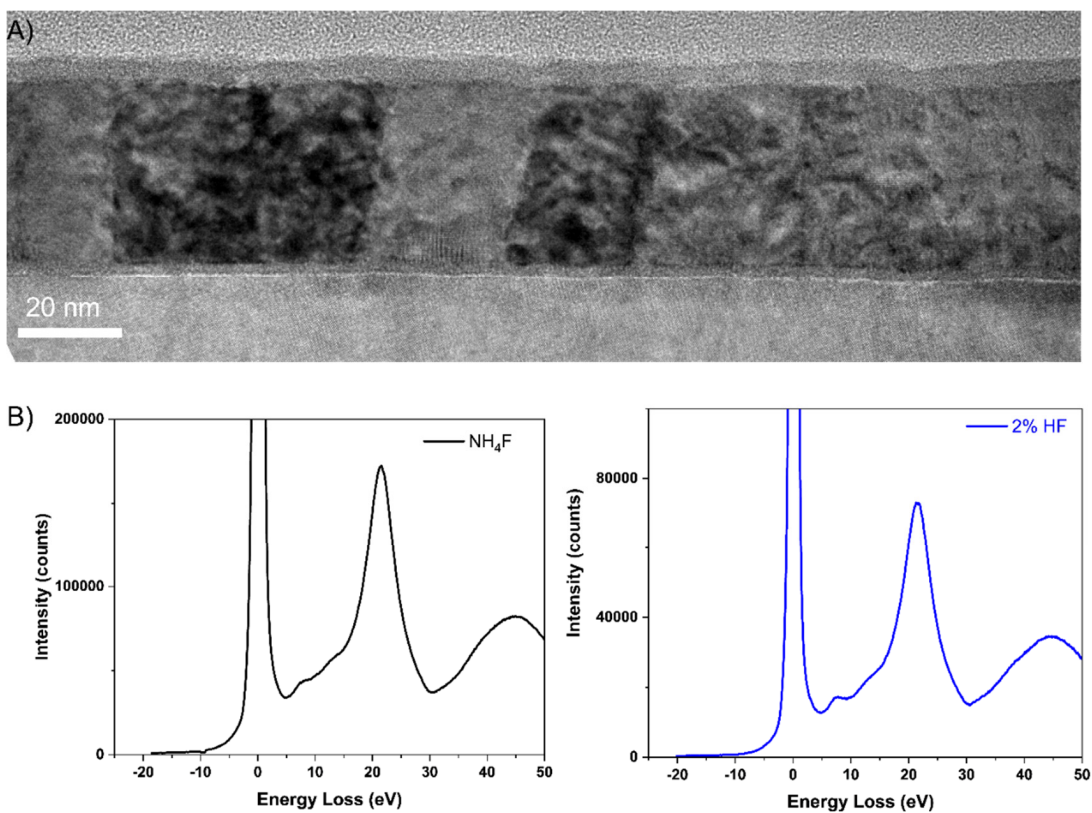

**Figure S7.** (A) Bright-field transmission electron microscopy (TEM) image of the NH$_4$F-Nb sample showing its columnar grain structure. (B) Raw low-loss electron energy loss spectroscopy (EELS) data for the NH$_4$F-Nb and 2% HF-Nb samples before background subtraction. A peak at 5-8 eV is apparent in both spectra. EELS characterization in the low-loss regime is qualitative, which we employ to further confirm the presence of Nb-H phase (bonding) in our treated samples. The plasmon peak intensity depends on several factors, such as the electron beam intensity and the sample thickness, which we did not monitor in our TEM experiments.

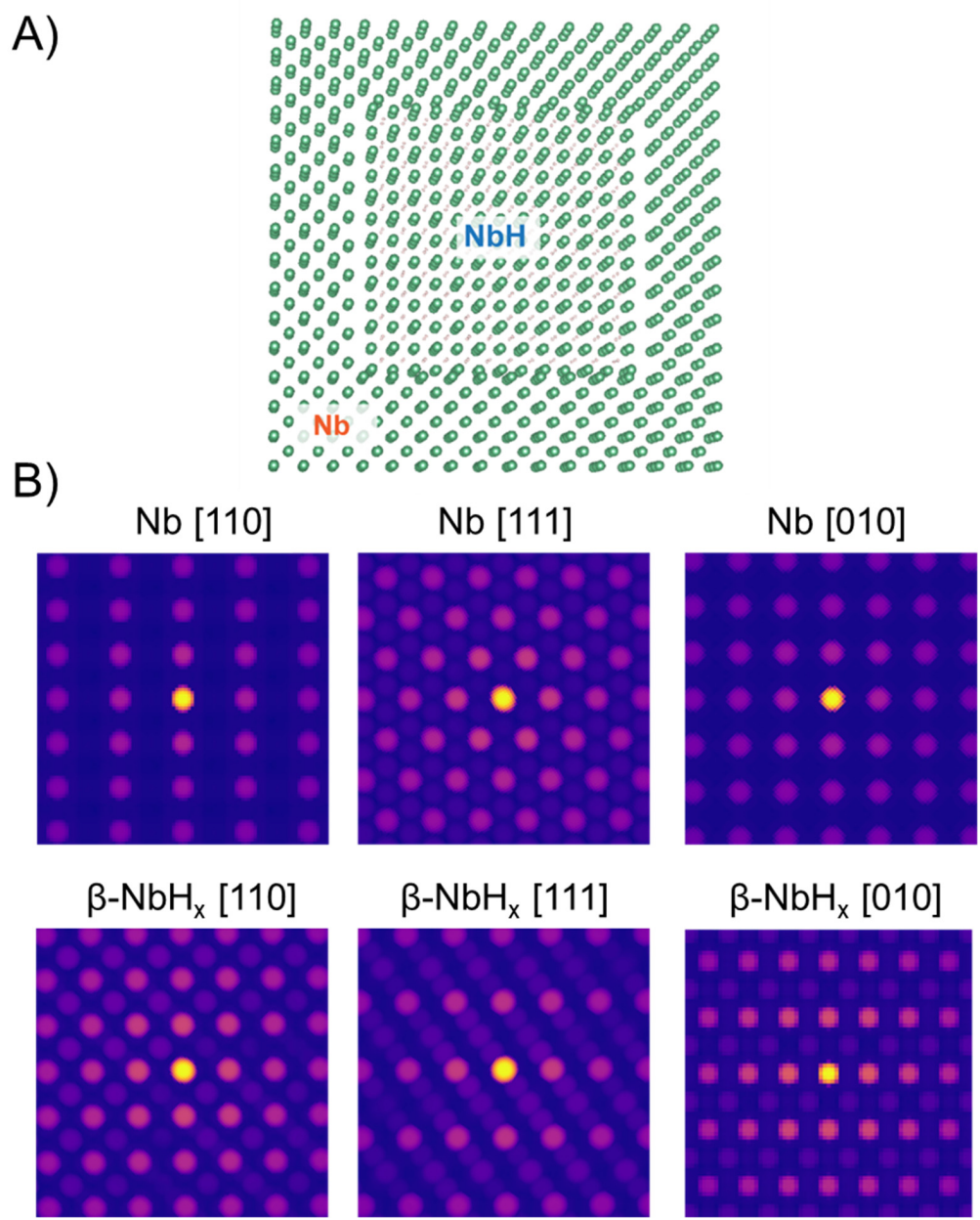

**Figure S8.** (A) Atomic model of a β-NbH$_x$ grain embedded within a Nb matrix. (B) Simulated CBED patterns (4D-STEM patterns) of Nb and β-NbH$_x$ at different orientations under conditions similar to the experimental microscopy parameters. (C) The same region of interest as shown in Fig. 3. The averaged CBED patterns of the orange and green areas are displayed in (D). The simulated patterns are shown side by side to confirm the two areas corresponding to Nb [020] (orange) and Nb [111] (green).

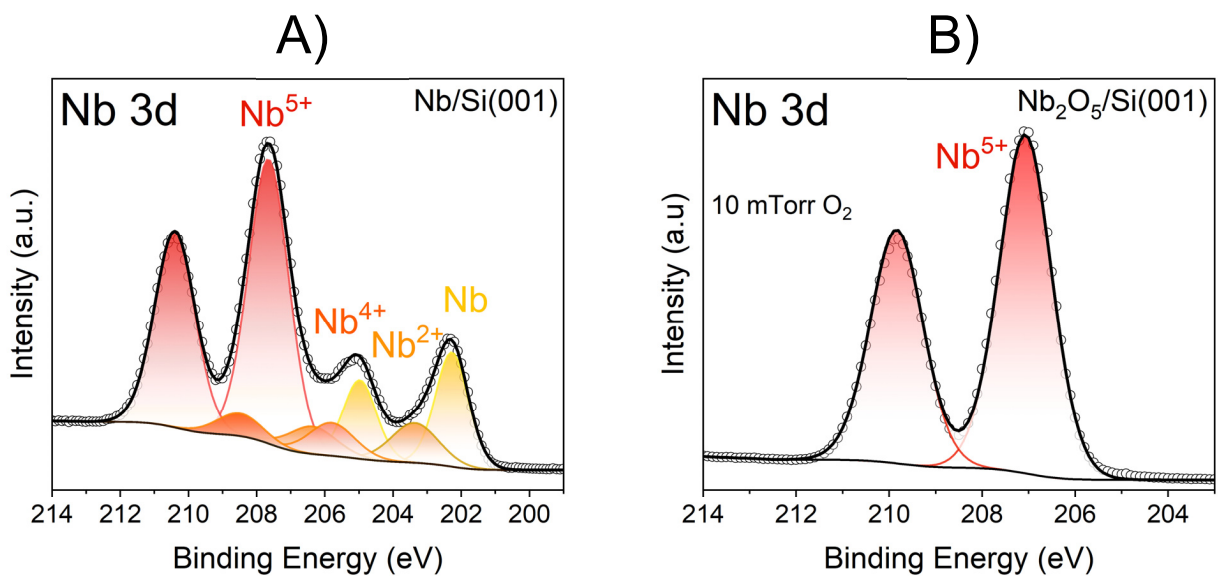

**Figure S9.** X-ray photoelectron spectroscopy (XPS) of the (A) Nb/Si(001) and (B) $Nb_2O_5$/Si(001) thin films. Both films are ~40 nm thick.

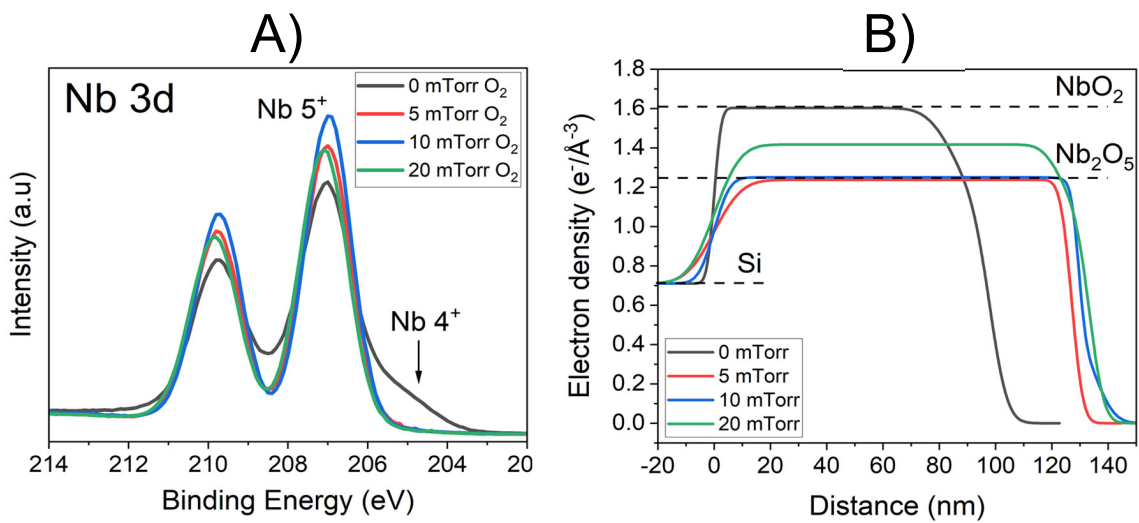

**Figure S10.** (A) XPS of $Nb_2O_5$ films deposited at different $O_2$ pressures. (B) Electron density profiles extracted from XRR fittings. The electron density of $Nb_2O_5$ and $NbO_2$ are shown as references.

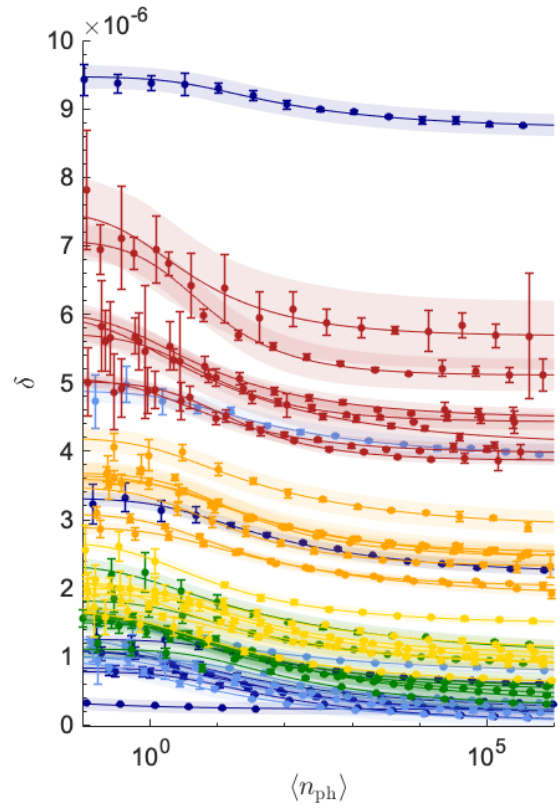

**Figure S11.** Dielectric loss as a function of the number of photons for Nb resonators after 20 min etching treatments: Control (light blue), $NH_4F$ (green), BOE (dark blue), 2% HF (yellow), 5% HF (orange), and 8% HF (red).

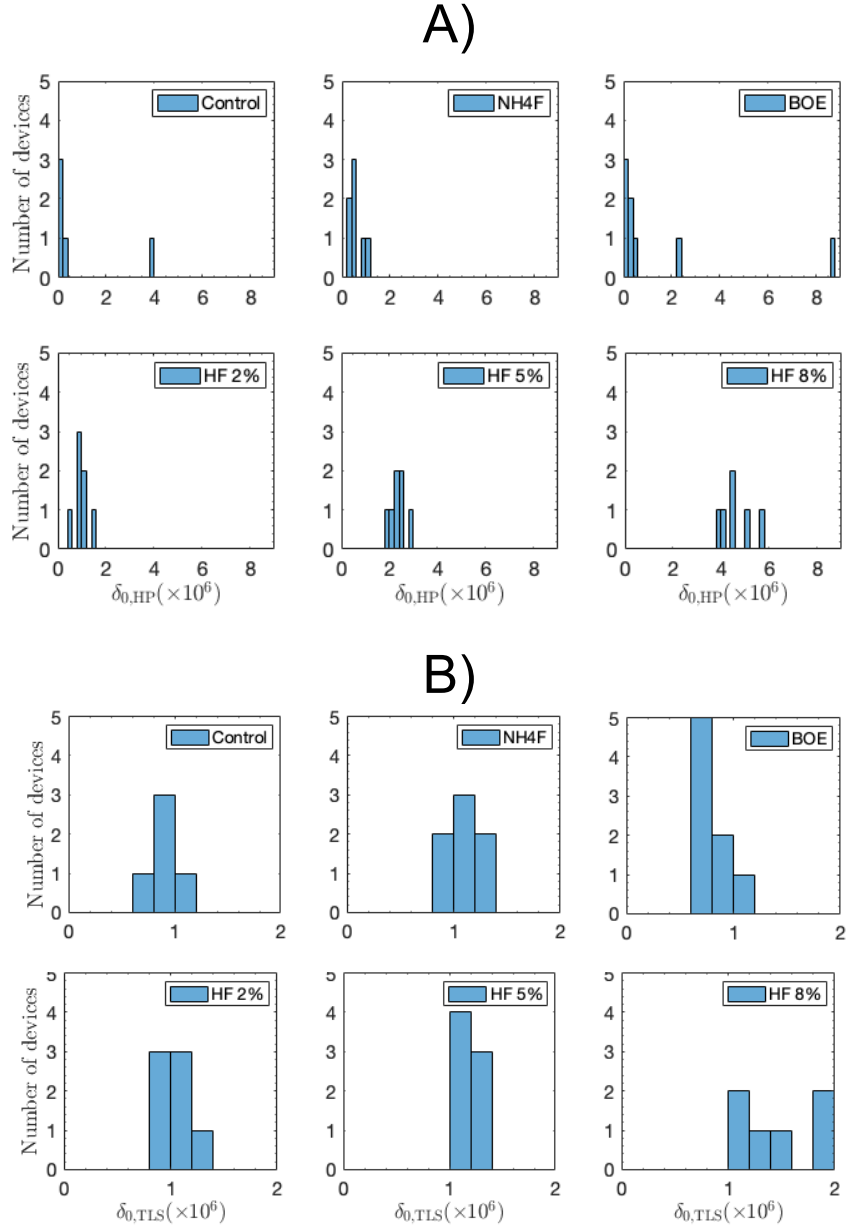

**Figure S12.** Histograms of (A) high power and (B) two-level system (TLS) loss extracted from the nonlinear least square fit to the TLS power-dependent model for the measured Nb resonators after 20 min wet etch treatments.

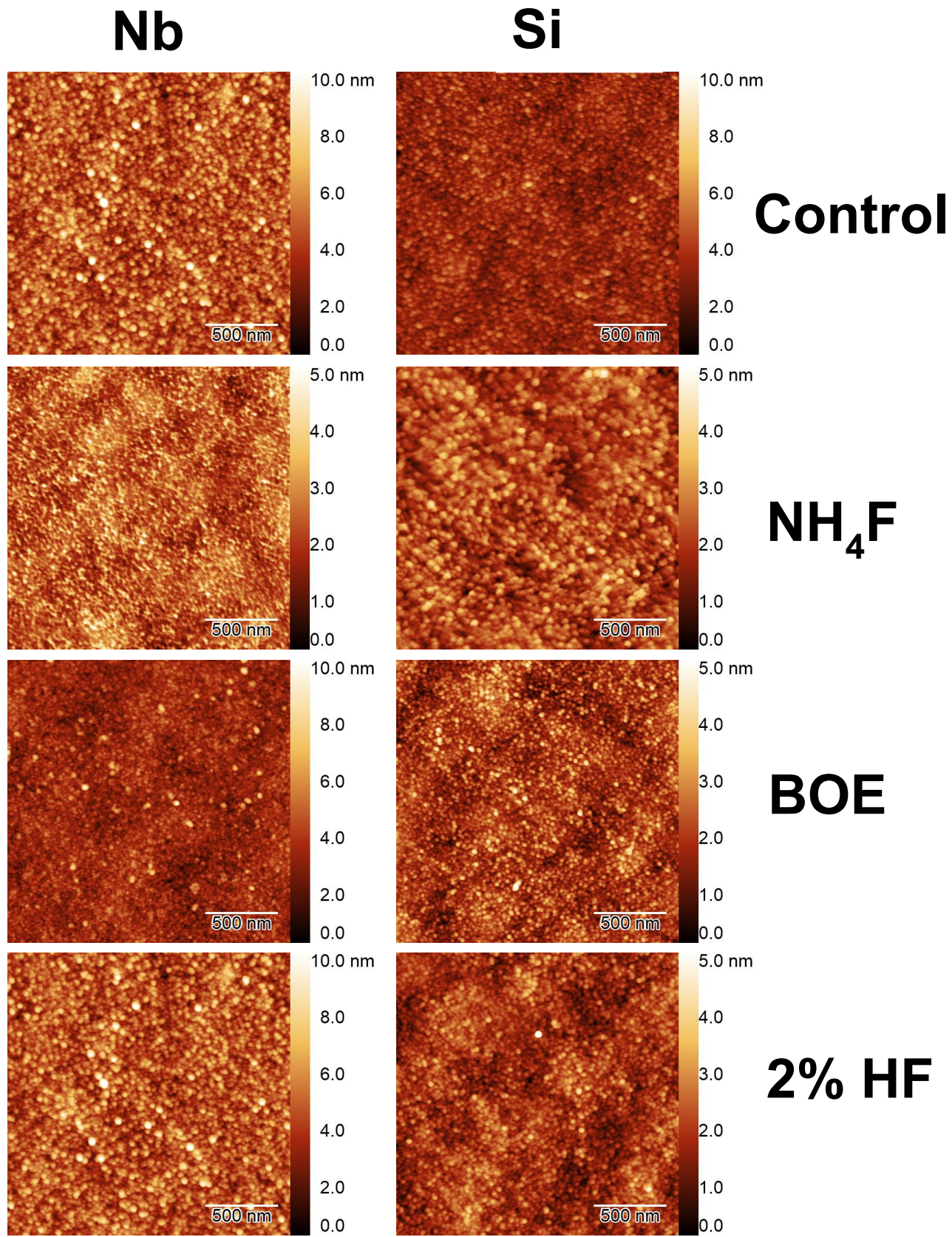

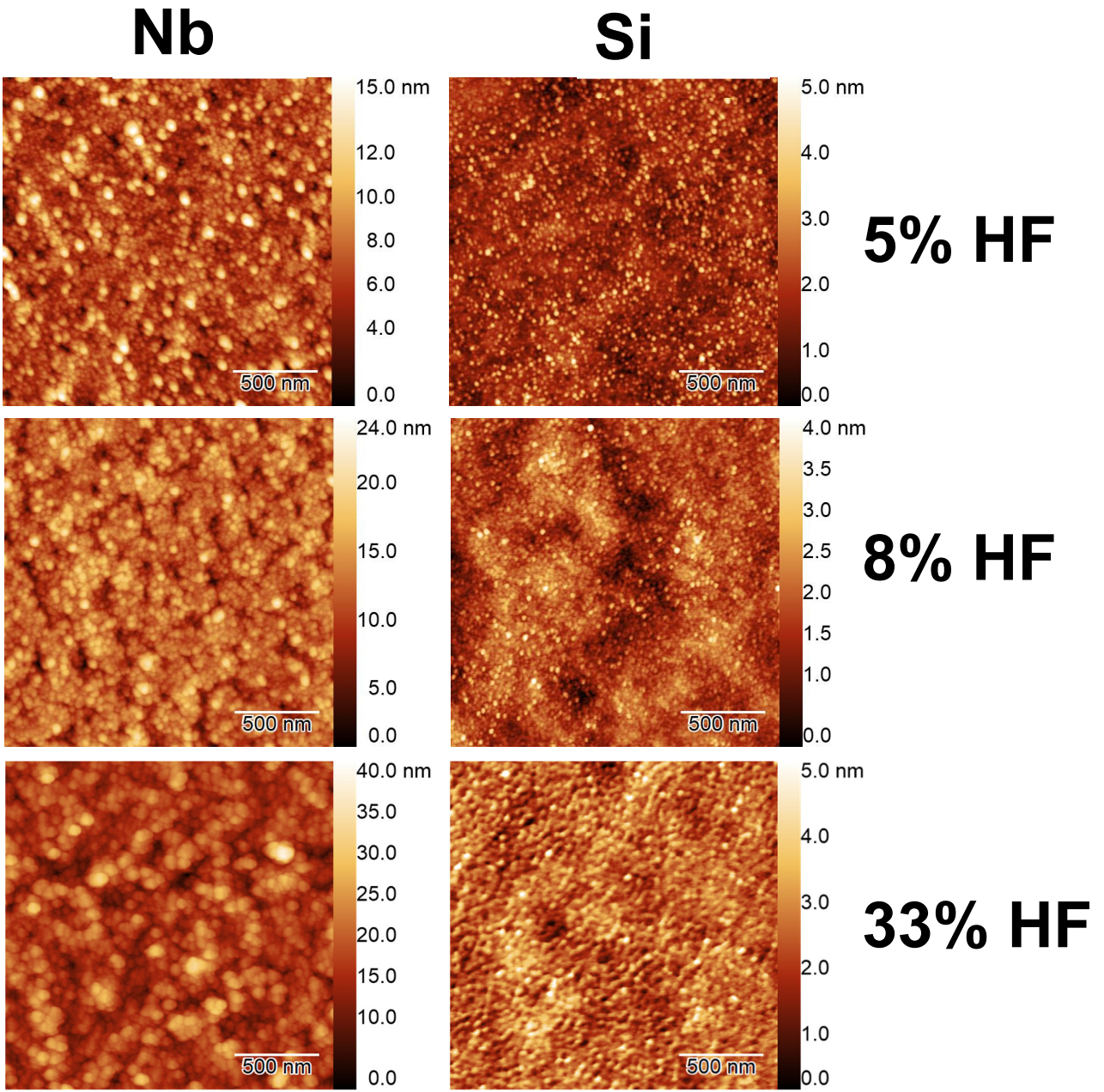

**Figure S13.** Atomic force microscopy (AFM) images of the Nb (left) and Si (right) locations in the resonators after 20 min etching treatment.

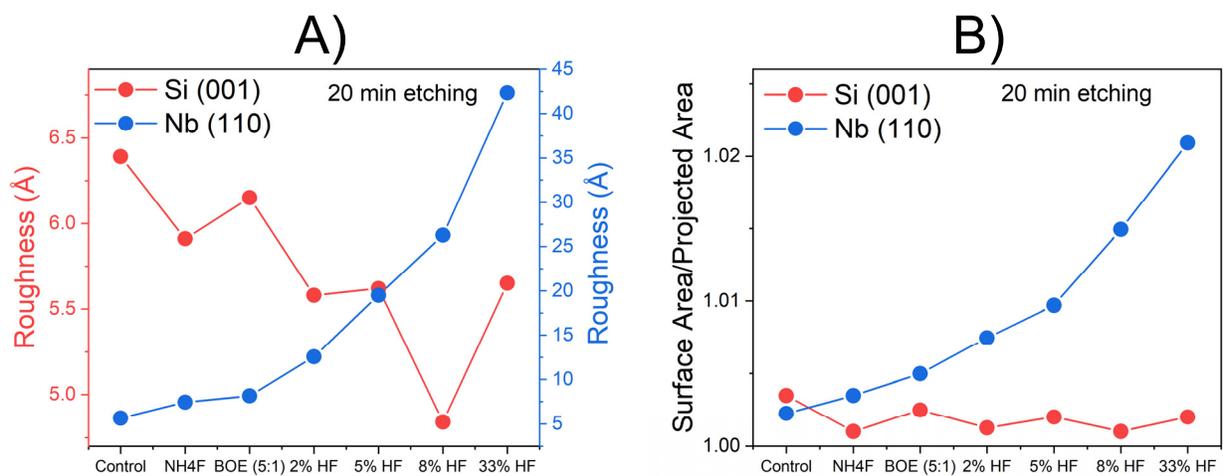

**Figure S14.** AFM-determined (A) RMS roughness and (B) projected surface area of the Nb and Si surfaces in the Nb resonators after 20 min etching treatments.

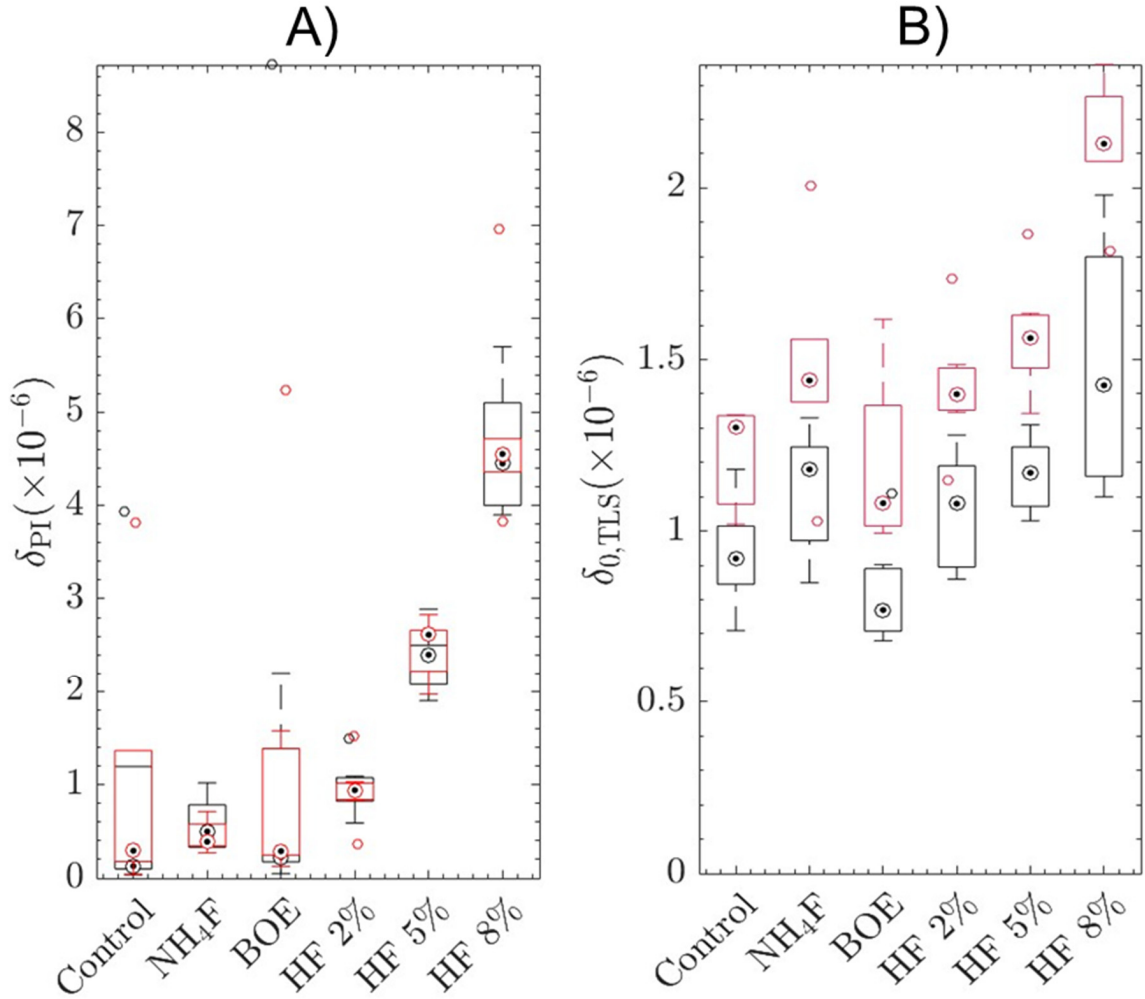

**Figure S15.** Effect of aging on (A) power independent and (B) two-level system (TLS) loss extracted from the nonlinear least square fit to the TLS power-dependent model for the Nb resonators. Black corresponds to the non-aged resonators and red to the aged resonators. The boxes contain the 25$^{th}$ and 75$^{th}$ percentiles, the full circles show the 95% confidence intervals about the median, the empty circles are the outliers, and the horizontal lines are the minimum and maximum values excluding outliers.